\documentclass[twocolumn,preprintnumbers,superscriptaddress,nofootinbib,aps,prd,floatfix]{revtex4-2}
\pdfoutput=1
\usepackage{enumerate}
\usepackage{amsmath,amssymb}
\usepackage{graphicx}
\usepackage{slashed}
\usepackage{xspace, slashed}
\usepackage[dvipsnames]{xcolor}
\usepackage{float}
\usepackage{hyperref}
\hypersetup{colorlinks=true, citecolor=Blue, urlcolor=Blue, linkcolor=Blue}
\usepackage[normalem]{ulem}
\usepackage{subfigure,orcidlink}
\usepackage{multirow,array}
\usepackage{tabularray}
\UseTblrLibrary{booktabs}
\usepackage{bbding}

\begin{document}	

\title{Impact of conversion-driven processes on singlet-doublet\\Majorana dark matter relic}


\author{Partha Kumar Paul \orcidlink{https://orcid.org/0000-0002-9107-5635}}
\email{ph22resch11012@iith.ac.in}
\affiliation{Department of Physics, Indian Institute of Technology Hyderabad, Kandi, Telangana-502285, India.}

\author{Sujit Kumar Sahoo \orcidlink{https://orcid.org/0000-0002-9014-933X}}
\email{ph21resch11008@iith.ac.in}
\affiliation{Department of Physics, Indian Institute of Technology Hyderabad, Kandi, Telangana-502285, India.}

\author{Narendra Sahu \orcidlink{https://orcid.org/0000-0002-9675-0484}}
\email{nsahu@phy.iith.ac.in}
\affiliation{Department of Physics, Indian Institute of Technology Hyderabad, Kandi, Telangana-502285, India.}
	
\date{\today}

\begin{abstract}
The singlet-doublet dark matter model offers a rich framework for exploring the nature of dark matter (DM) through its unique fermion structure. In this model, the important parameters are the singlet-doublet mass splitting $\Delta{M}$, the singlet-doublet mixing angle $\sin\theta$, and the DM mass $M_{\rm DM}$. If the DM is assumed to be of Dirac nature, then the annihilation, co-annihilation, and conversion driven processes combined allow a range of parameter space: $100~{\rm GeV} \lesssim M_{\rm DM}\lesssim750$ GeV and $10^{-6}\lesssim\sin\theta\lesssim0.04$ for $\Delta{M}>1$ GeV. While the nature of DM, either Dirac or Majorana, is not known, in this work, we assume the singlet-doublet DM to be of Majorana type and find that the relic density and direct detection can be satisfied over a larger parameter space. In particular, the allowed ranges of DM mass and $\sin\theta$ are $100~{\rm GeV}\lesssim M_{\rm DM}\lesssim1750$ GeV and $2\times10^{-7}\lesssim\sin\theta\lesssim0.45$ for $\Delta{M}>1$ GeV. 
\end{abstract}	

\preprint{\href{https://doi.org/10.1103/kctx-7kfl}{Phys. Rev. D \textbf{113} (2026) 9, 095040}}

\maketitle

\section{Introduction}\label{intro}

Dark matter (DM) constitutes approximately 26.8\% of the Universe's total energy budget ($\Omega_{\rm DM}h^2\simeq 0.12$) \cite{Planck:2018vyg}, evidencing its presence through galaxy rotation curves, gravitational lensing, the formation of large-scale structures, etc \cite{Zwicky:1933gu,Rubin:1970zza}. Given the absence of a DM candidate within the standard model (SM), the weakly interacting massive particle (WIMP) emerges as a compelling contender, capable of accounting for the DM relic density through freeze-out processes in the early Universe. A simple extension of the SM for the explanation of DM could be a singlet fermion, which is odd under $\mathcal{Z}_2$ symmetry. However, the parameter space faces stringent constraints from both relic density requirements and direct detection experiments \cite{Bhattacharya:2018fus,Dutta:2020xwn}. Another alternative would be to consider a $\mathcal{Z}_2$ odd doublet fermion, which also fails due to excessive annihilation and detection constraints. These issues can be resolved if one considers the combination of singlet and doublet fermions. The neutral component of the doublet mixes with the singlet fermion to give the so-called fermionic singlet-doublet DM (SDDM) \cite{Cynolter:2015sua, Bhattacharya:2015qpa, Bhattacharya:2018fus, Bhattacharya:2017sml, Bhattacharya:2018cgx, Bhattacharya:2016rqj, Dutta:2020xwn, Borah:2021khc, Borah:2021rbx, Borah:2022zim, Borah:2023dhk, Paul:2024iie, Mahbubani:2005pt, DEramo:2007anh, Cohen:2011ec, Freitas:2015hsa, Calibbi:2015nha, Cheung:2013dua, Banerjee:2016hsk, DuttaBanik:2018emv, Horiuchi:2016tqw, Restrepo:2015ura, Abe:2017glm, Konar:2020wvl, Konar:2020vuu, Calibbi:2018fqf, Ghosh:2021wrk, Das:2023owa, Bhattacharya:2021ltd, Enberg:2007rp,Paul:2024prs}. The attractive part of this model is that it is a very minimal model in the beyond the SM scenario, having only three parameters: DM mass $M_{\rm DM}$, singlet-doublet (SD) mixing $\sin\theta$, and SD mass splitting $\Delta{M}$. This model not only predicts the DM relic but also facilitates probes at different terrestrial experiments. The DM relic is decided not only by annihilation and co-annihilation processes but also by conversion-driven \cite{DAgnolo:2017dbv,Garny:2017rxs,Maity:2019hre} processes (which include co-scattering, decay, and inverse-decay). It was found that, if the SD mixing is large (typically $\sin\theta\gtrsim0.05$), annihilation and co-annihilation processes determine the relic abundance \cite{Bhattacharya:2018fus}. On the other hand, in the small-mixing-angle limit, conversion-driven processes play an important role in determining the relic of DM. It was shown that for $\sin\theta\gtrsim10^{-6}$ the singlet can be brought to equilibrium \cite{Paul:2024prs}. However, for small $\sin\theta$, the singlet component decouples early with a larger abundance. Nevertheless, through conversion-driven processes, the DM particles scatter into a higher-mass state within the dark sector, such as the doublet, which then annihilates into SM particles, thereby reducing the singlet abundance. As a result, the final relic of the DM can be brought to the correct ballpark even if $\sin\theta$ is small. 
For $\sin\theta<10^{-7}$, the singlet never attains equilibrium. In this case, the relic can be produced by non-thermal mechanisms, such as SuperWIMP and freeze-in.  Assuming the SDDM to be Dirac type, the full parameter space of the  DM relic, including annihilation, co-annihilation, co-scattering, decay, and inverse decay, was estimated in \cite{Paul:2024prs}.  In fact, the allowed parameter space for a Dirac SDDM was found to be  $100~{\rm GeV} \lesssim M_{\rm DM}\lesssim750$ GeV and $10^{-6}\lesssim\sin\theta\lesssim0.04$ for $\Delta{M}>1$ GeV. On the other hand, in the Majorana fermionic DM scenario, where the DM particle is its own antiparticle, its dynamics or the relic-determining processes differ from those of its Dirac counterpart. Consequently, this distinction leads to a modified parameter space for Majorana DM candidates.

In this paper, we assume the SDDM to be of Majorana type and study the allowed parameter space for the correct relic density, including annihilation, co-annihilation, and conversion-driven processes. We show that in the case of Majorana SDDM, the correct relic density allows the ranges of DM mass and $\sin\theta$ are: $100~{\rm GeV}\lesssim M_{\rm DM}\lesssim1750$ GeV and $2\times10^{-7}\lesssim\sin\theta\lesssim0.45$ for $\Delta{M}>1$ GeV. We also note that inclusion of the conversion-driven processes shifts the relic parameter space to the sensitivity ranges of LHC \cite{ATLAS:2022gbw,CMS:2024trg}  and MATHUSLA \cite{MATHUSLA:2019qpy}.

The paper is organized as follows. In Sec. \ref{sec:model}, we describe the singlet-doublet Majorana DM model in detail. The DM phenomenology and detection prospects are discussed in Sec. \ref{sec:dmpheno}. We finally conclude in Sec. \ref{sec:conc}.

\section{The Model}\label{sec:model}
The SM is extended with a singlet fermion, $\chi$, and a doublet fermion, $\Psi=(\psi^0~~\psi^-)^T=(\psi^0_L+\psi^0_R~~~\psi^-)^T$, which are odd under a discrete symmetry $\mathcal{Z}_2$. The DM Lagrangian can be written as

\begin{eqnarray}
	\mathcal{L}_{\rm DM}&=&\overline{\Psi}i\gamma^\mu\mathcal{D}_\mu\Psi-M_{\Psi}\bar{\Psi}\Psi+\overline{\chi}i\gamma^\mu\partial_\mu \chi-\frac{1}{2}M_\chi\overline{\chi^c}\chi\nonumber\\&&
	-\frac{y_\chi}{\sqrt{2}}\overline{\Psi}\tilde{H}(\chi+\chi^c)+{\rm H.c.}.\nonumber\\
\end{eqnarray}

The neutral fermion mass matrix can be written in the basis $((\psi^0_R)^c,\psi^0_L,(\chi)^c)^T$ as

\begin{equation}
\begin{pmatrix}
	0 & M_\Psi & \frac{m_D}{\sqrt{2}}\\
	M_\Psi&0&  \frac{m_D}{\sqrt{2}}\\
	 \frac{m_D}{\sqrt{2}} &  \frac{m_D}{\sqrt{2}}& M_\chi
\end{pmatrix},
\end{equation}
where $m_D=y_\chi v_h/\sqrt{2}$ and $v_h$ is the vacuum expectation value (vev) of the SM Higgs boson, $H$. The mass matrix can be diagonalized with a unitary matrix of the form $U(\theta)=P.U_{13}(\theta_{13}=\theta).U_{23}(\theta_{23}=0).U_{12}(\theta_{12}=\frac{\pi}{4})$. The details can be found in Appendix \ref{app:massdiag}.
The three neutral states mix and give three Majorana states as $\chi_i=\frac{\chi_{iL}+\chi_{iL}^c}{\sqrt{2}}$, where
\begin{eqnarray}
	\chi_{1L}&=&\frac{\cos\theta}{\sqrt{2}}(\psi^0_L+(\psi^0_R)^c)+\sin\theta N_1^c,\nonumber\\
\chi_{2L}&=&\frac{i}{\sqrt{2}}(\psi^0_L-(\psi^0_R)^c),\nonumber\\
\chi_{3L}&=&-\frac{\sin\theta}{\sqrt{2}}(\psi^0_L+(\psi^0_R)^c)+\cos\theta N_1^c.
\end{eqnarray}
The corresponding mass eigenvalues are
\begin{eqnarray}\label{eq:massbasis}
M_{\chi_1}&=&M_\Psi\cos^2\theta+M_\chi\sin^2\theta+m_D\sin2\theta,\nonumber\\M_{\chi_2}&=&M_\Psi,\nonumber\\
M_{\chi_3}&=&M_\Psi\sin^2\theta+M_\chi\cos^2\theta-m_D\sin2\theta,
\end{eqnarray}
where the mixing angle is given as
\begin{eqnarray}\label{eq:theta_deltaM}
	\tan2\theta=\frac{2m_D}{M_\Psi-M_{\chi}}.
\end{eqnarray}
In Appendix \ref{app:lagMass}, we have provided the interaction Lagrangian in the mass basis. In this setup, we consider $M_{\chi_3}$ to be the lightest among the dark sector particles and identify $\chi_3$ as the DM candidate with $M_{\chi_3}=M_{\rm DM}$. The Yukawa coupling, using Eq. \ref{eq:sin2th}, can be expressed as
\begin{eqnarray}
y_\chi=\frac{{\Delta}M\sin2\theta}{\sqrt{2}v_h},
\end{eqnarray}
where the SD mass splitting $\Delta M=M_{\chi_1}-M_{\chi_3}$.
\section{dark matter phenomenology and detection prospects}\label{sec:dmpheno}
We investigate the DM phenomenology of the SDDM model, considering a singlet Majorana fermion as the DM candidate within a thermal freeze-out framework. A comprehensive numerical scan is performed over the model parameter space, incorporating conversion-driven processes alongside conventional annihilation and co-annihilation mechanisms (respective Feynman diagrams are provided in Appendix~\ref{app:FD}). The Feynman diagrams relevant for the relic density calculation are given in Appendix \ref{app:FD}. The analysis explores the viable mass range for the DM candidate between 10 and 2000 GeV. In earlier studies of the SDDM framework \cite{Dutta:2020xwn}, it was commonly assumed that all dark sector particles decouple simultaneously. However, this assumption does not necessarily hold for small mixing angles. In particular, for values of $\sin\theta \lesssim 0.05$, the DM particles fail to maintain chemical equilibrium with the thermal bath. Consequently, the singlet typically decouples earlier, while the doublet remains in equilibrium for a longer period owing to its gauge interactions.

In accordance with this, we consider three distinct sectors to trace the evolution of DM independently of other particles in the thermal bath. The DM component ($\chi_3$) is assigned to sector 1, while the remaining dark sector particles ($\chi_1,\chi_2,\psi^\pm$) belong to sector 2, and the SM particles constitute sector 0. Note that the fermion doublet components ($\chi_1,\chi_2,\psi^\pm$) maintain chemical equilibrium among themselves with the help of gauge-mediated interactions, e.g., processes like $\psi^\pm\chi_{1,2}\leftrightarrow{\rm SM}~\rm SM$ with $W^\pm$ exchange and that decouple at the same epoch. This justifies treating them as a single sector, i.e., sector 2. In this setup, particles in sectors 1 and 2 can interact with each other and with sector 0 (i.e., SM particles). Consequently, the cosmological evolution of sectors 1 and 2 must be tracked simultaneously for all ranges of $\sin\theta$. We define the comoving number densities of sector 1 and sector 2, as $Y_{1}(=n_{\chi_3}/s)$ and $Y_{2}\left(=(n_{\chi_1}+n_{\chi_2}+n_{\psi^\pm})/s\right)$, respectively, where $n_i$ denotes the number density of the $i$th species and $s=2\pi^2g_{*s}T^3/45$ represents the entropy density. The Boltzmann equations governing the evolution of $Y_1$ and $Y_2$ are given by \cite{Paul:2024prs,Alguero:2022inz},
\begin{widetext}
\begin{eqnarray}
		\frac{dY_1}{dT} &=&   \frac{1}{3\mathcal{H}}\frac{ds}{dT} \left[    \langle \sigma_{1100} v \rangle ( Y_1^2 - {Y_1^{\rm eq}}^2) +    \langle \sigma_{1122} v \rangle \left( Y_1^2 - Y_2^2  \frac{{Y_1^{\rm eq}}^2}{{Y_2^{\rm eq}}^2}\right)  + \langle \sigma_{1200} v \rangle ( Y_1 Y_2 - Y_1^{\rm eq}Y_2^{\rm eq})\right. \nonumber\\
		&&+\left.  \langle \sigma_{1222} v \rangle \left( Y_1 Y_2 - Y_2^2   \frac{Y_1^{\rm eq}}{Y_2^{\rm eq}} \right) -\langle \sigma_{1211} v \rangle \left( Y_1 Y_2 - Y_1^2   \frac{Y_2^{\rm eq}}{Y_1^{\rm eq}} \right)
		-\frac{ \Gamma_{2\rightarrow 1}}{s}\left( Y_2 -Y_1 \frac{Y_2^{\rm eq}}{Y_1^{\rm eq}}  \right)        \right] ,
		\label{eq:Y1}
	\end{eqnarray}
	\begin{eqnarray}
		\frac{dY_2}{dT} &=&   \frac{1}{3\mathcal{H}}\frac{ds}{dT}\left[    \langle \sigma_{2200} v \rangle ( Y_2^2 - {Y_2^{\rm eq}}^2) -    \langle \sigma_{1122} v \rangle \left( Y_1^2 - Y_2^2  \frac{{Y_1^{\rm eq}}^2}{{Y_2^{\rm eq}}^2}\right) +  \langle \sigma_{1200} v \rangle ( Y_1 Y_2 - Y_1^{\rm eq}Y_2^{\rm eq}) \right. \nonumber \\
		&&- \left. \langle \sigma_{1222} v \rangle \left( Y_1 Y_2 - Y_2^2   \frac{Y_1^{\rm eq}}{Y_2^{\rm eq}} \right)
		+\langle \sigma_{1211} v \rangle \left( Y_1 Y_2 - Y_1^2   \frac{Y_2^{\rm eq}}{Y_1^{\rm eq}} \right)  + \frac{ \Gamma_{2\rightarrow 1}}{s}\left( Y_2 -Y_1 \frac{Y_2^{\rm eq}}{Y_1^{\rm eq}}  \right)        \right],
		\label{eq:Y2}
	\end{eqnarray}
\end{widetext}
where $Y_i^{\rm eq}\left(=n_i^{\rm eq}/s\right)$ are the equilibrium abundances, $\mathcal{H}=1.66\sqrt{g_*}\frac{T^2}{M_{\rm Pl}}$ is the  Hubble parameter with $M_{\rm pl}=1.22\times10^{19}$ GeV being the Planck mass, and $\langle \sigma_{\alpha\beta\gamma\delta} v\rangle$ are the thermally averaged cross-sections for the processes $\alpha\beta\leftrightarrow\gamma\delta$. The term $\Gamma_{2\rightarrow1}$ in Eq. (\ref{eq:Y2}) includes both decay, $\Gamma_{\rm D}$, and co-scattering rates as given by
\begin{eqnarray}
    \Gamma_{2\rightarrow1}=\Gamma_{\rm D}\frac{K_1(M_{\Psi}/T)}{K_2(M_{\Psi}/T)}+n^{\rm eq}_{\rm SM}\langle\sigma_{2010}v\rangle.
\end{eqnarray}
\begin{figure*}[tbh]
    \centering
    \includegraphics[scale=0.4]{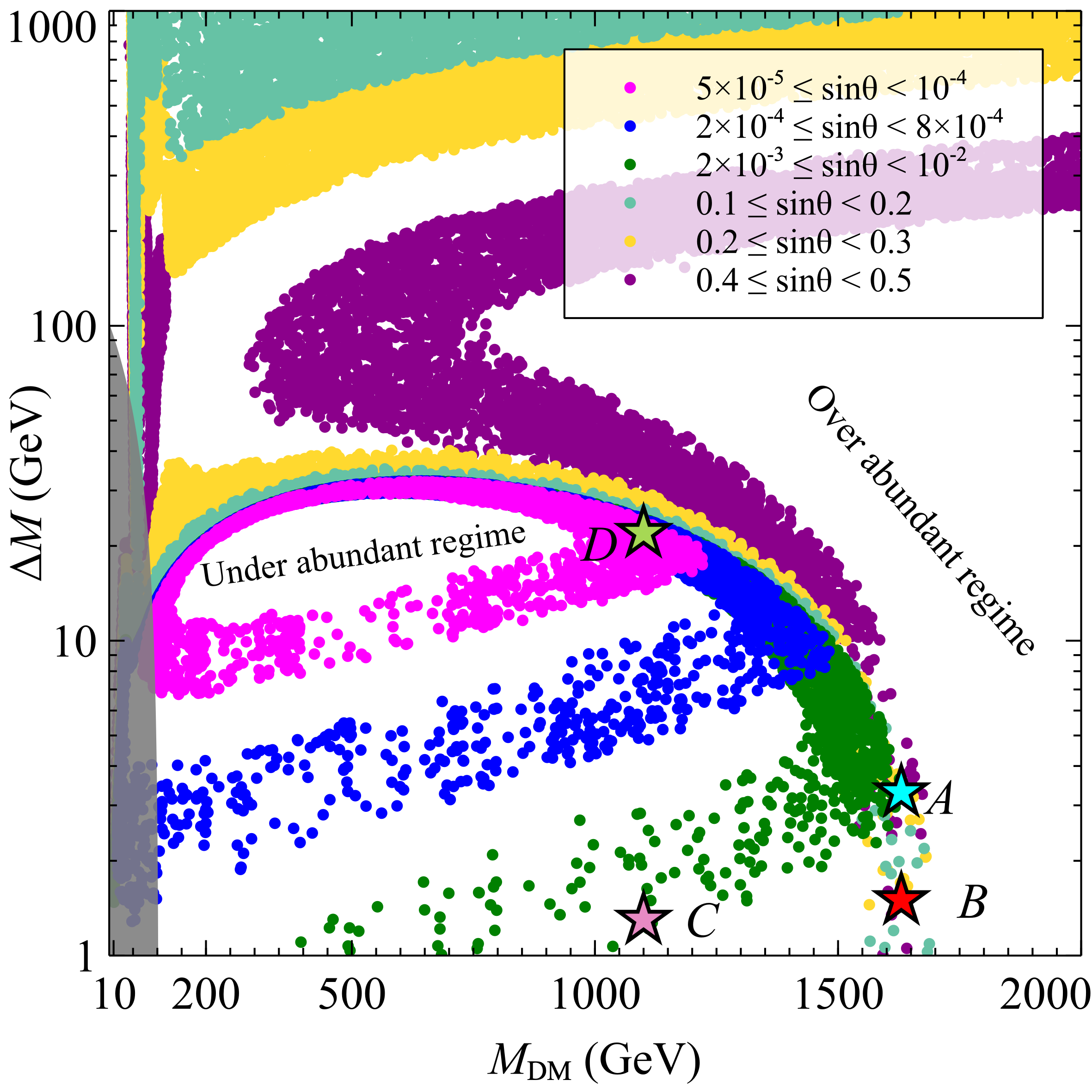}
    \includegraphics[scale=0.4]{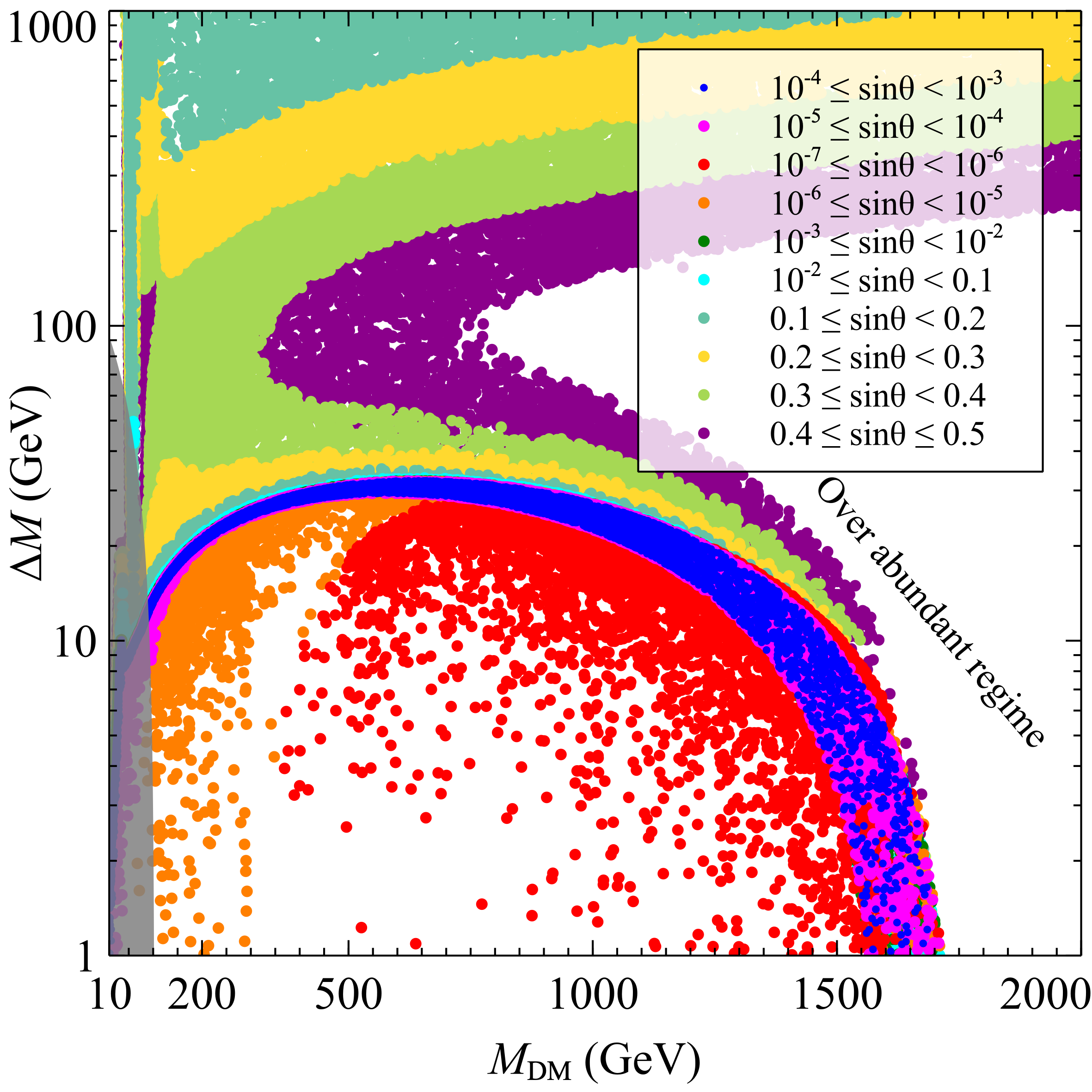}
    \caption{[\textit{Left:}] correct dark matter relic points are shown in the plane of $\Delta{M}$ vs $M_{\rm DM}$ for different ranges of $\sin\theta$ as mentioned in the inset of the figure, without considering co-scattering contributions while solving Eqs. (\ref{eq:Y1}) and (\ref{eq:Y2}). [\textit{Right:}] correct dark matter relic points are shown in the plane of $\Delta{M}$ vs $M_{\rm DM}$ for different ranges of $\sin\theta$ as mentioned in the inset of the figure, considering all the relevant contributions including annihilation, co-annihilation, co-scattering and decay, inverse-decay while solving Eqs. (\ref{eq:Y1}) and (\ref{eq:Y2}). The gray region is excluded from the LEP bound \cite{DELPHI:2003uqw}.}
    \label{fig:delmvsmdm}
\end{figure*}
In the \textit{left} panel of Fig.~\ref{fig:delmvsmdm}, we present the parameter space yielding the observed relic abundance in the $\Delta M$–$M_{\rm DM}$ plane, obtained by numerically solving Eqs.~(\ref{eq:Y1}) and (\ref{eq:Y2}) without incorporating the co-scattering contributions in the $\Gamma_{2\rightarrow1}$ term\footnote{We use \texttt{micrOMEGAs} \cite{Alguero:2023zol} to calculate the relic density. Here we explicitly verify the thermalization of the DM by considering $\Gamma/\mathcal{H}>1$ using the function \texttt{checkTE}, where $\Gamma$ is the total interaction rate and $\mathcal{H}$ is the Hubble parameter.}. The different colors correspond to distinct ranges of the SD mixing angle $\sin\theta$, as indicated in the figure inset. At larger mass splittings ($\Delta M \gtrsim 100~\text{GeV}$), DM interacts with the SM bath via annihilation processes, providing a correct freeze-out relic, while for smaller $\Delta{M}$ ($\Delta M \lesssim 100~\text{GeV}$), the co-annihilation processes become dominant in comparison to the annihilation processes. Nevertheless, the decay and inverse-decay processes also play a significant role in maintaining equilibrium among the sector 1 and sector 2 particles \cite{Paul:2024prs}.
As the DM mass increases, the cross-section decreases. To enhance the cross-section in the annihilation-dominant region, $\Delta{M}$ has to be increased to get the relic abundance of DM in the correct ballpark. For the same reason, in the co-annihilation dominating region, $\Delta M$ has to be reduced with increasing $M_{\rm DM}$ to obtain the correct relic density. This is why, for $\sin\theta \gtrsim 10^{-1}$, we see an upward trend in $\Delta M$ in the region $\Delta M\gtrsim 100$ GeV, while a downward trend in $\Delta M$ for $\Delta M\lesssim 100$ GeV with an increase in DM mass. This leads to a minimum value of $M_{\rm DM}$ for each $\sin\theta$ as shown by different colored bands.  For each color band, the region to its right corresponds to an overabundant relic density, whereas the region to its left leads to an underabundant relic density. In the small $\sin\theta~(\lesssim0.05)$ region, the dominating processes that decide the relic density are co-annihilation, decay, and inverse-decay. In this region, due to the interplay between co-annihilation and decay and inverse-decay processes\footnote{Note that the interaction rate of the co-annihilation process enhances with a decrease in $\Delta M$, while the decay width decreases with decrease in $\Delta M$.}, there exists a maximum allowed $M_{\rm DM}$ for each $\sin\theta$ beyond which the relic remains overabundant.
To understand this behavior, we have provided four benchmark points, namely, $A, B, C$, and $D$ in Table \ref{tab:tab1}.
\begin{table*}[tbh]
		\centering		\caption{Relic density for benchmark points are given: (a)  without considering co-scattering, and (b) without considering co-scattering and decay\footnote{We get the relic ``$\Omega_{2s}h^2$(no co-scattering) without decay" by using a flag \texttt{ExcludedForNDM="2010 DMdecay"} inside \texttt{micrOMEGAs}. On the other hand ``$\Omega_{2s}h^2$(no co-scattering) with decay" is obtained by using \texttt{ExcludedForNDM="2010"}}.}
			\begin{tabular}{|c|c|c|c|c|c|}
				\hline\hline
				Point & $M_{\rm DM}$ (GeV) &$\Delta{M}$ (GeV) & $\sin\theta$ & $\Omega_{2s}h^2$(no co-scattering) with decay& $\Omega_{2s}h^2$(no co-scattering) without decay\\
				\hline
				$A$&1630&3.3 &$9.5\times10^{-3}$ & 0.12&0.896\\
                \hline
				$B$&1630&1.5 &$9.5\times10^{-3}$ & 0.21&$0.796$\\
                \hline
				$C$&1100&1.3 &$9.5\times10^{-3}$ & 0.12&$1.21$\\
                \hline
				$D$&1100&22&$9.5\times10^{-3}$ & 0.12&$5.1$\\
				\hline\hline
			\end{tabular}
		\label{tab:tab1}
\end{table*}
\begin{figure*}[tbh]
    \centering
\includegraphics[scale=0.5]{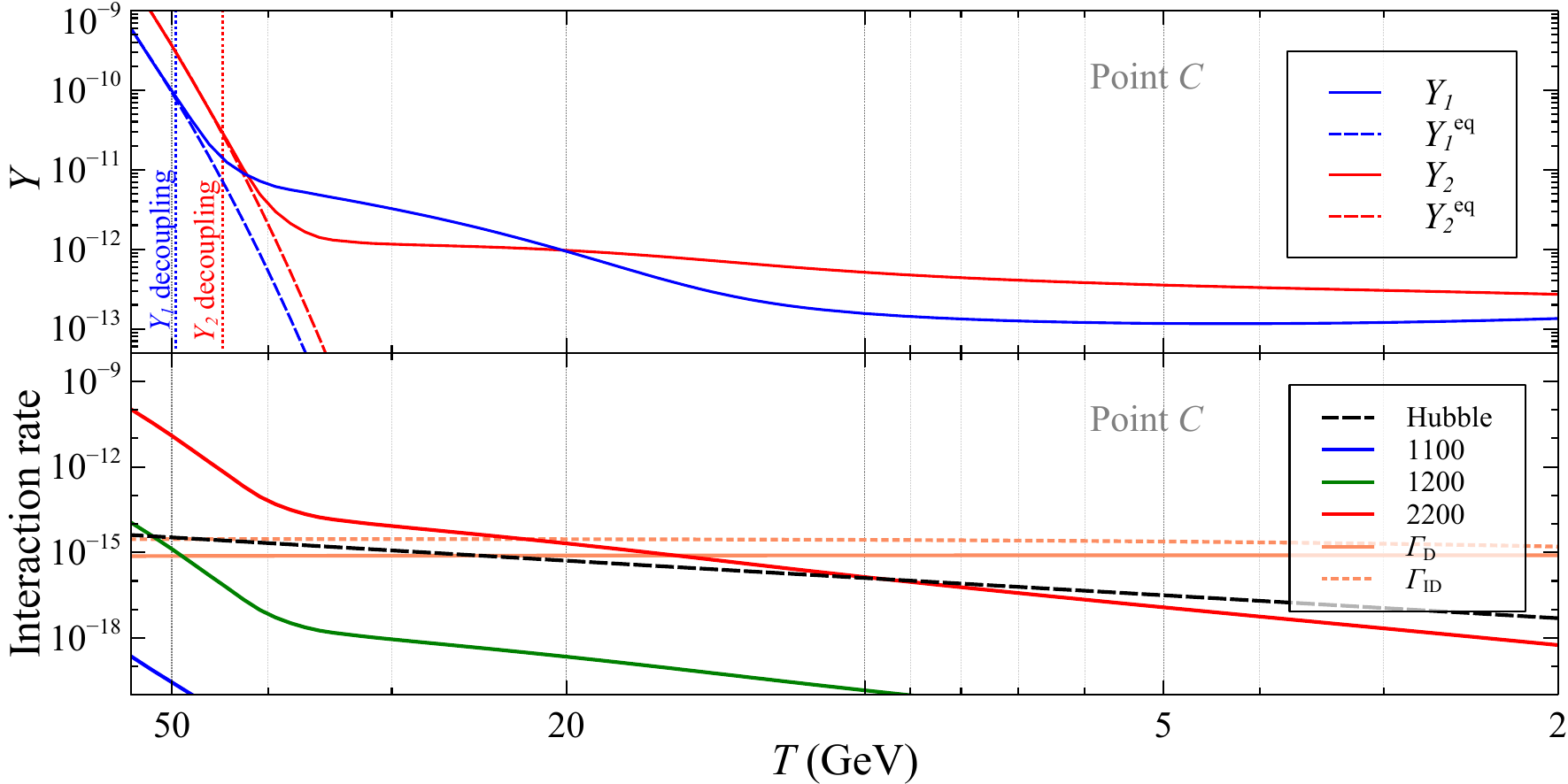}
\includegraphics[scale=0.5]{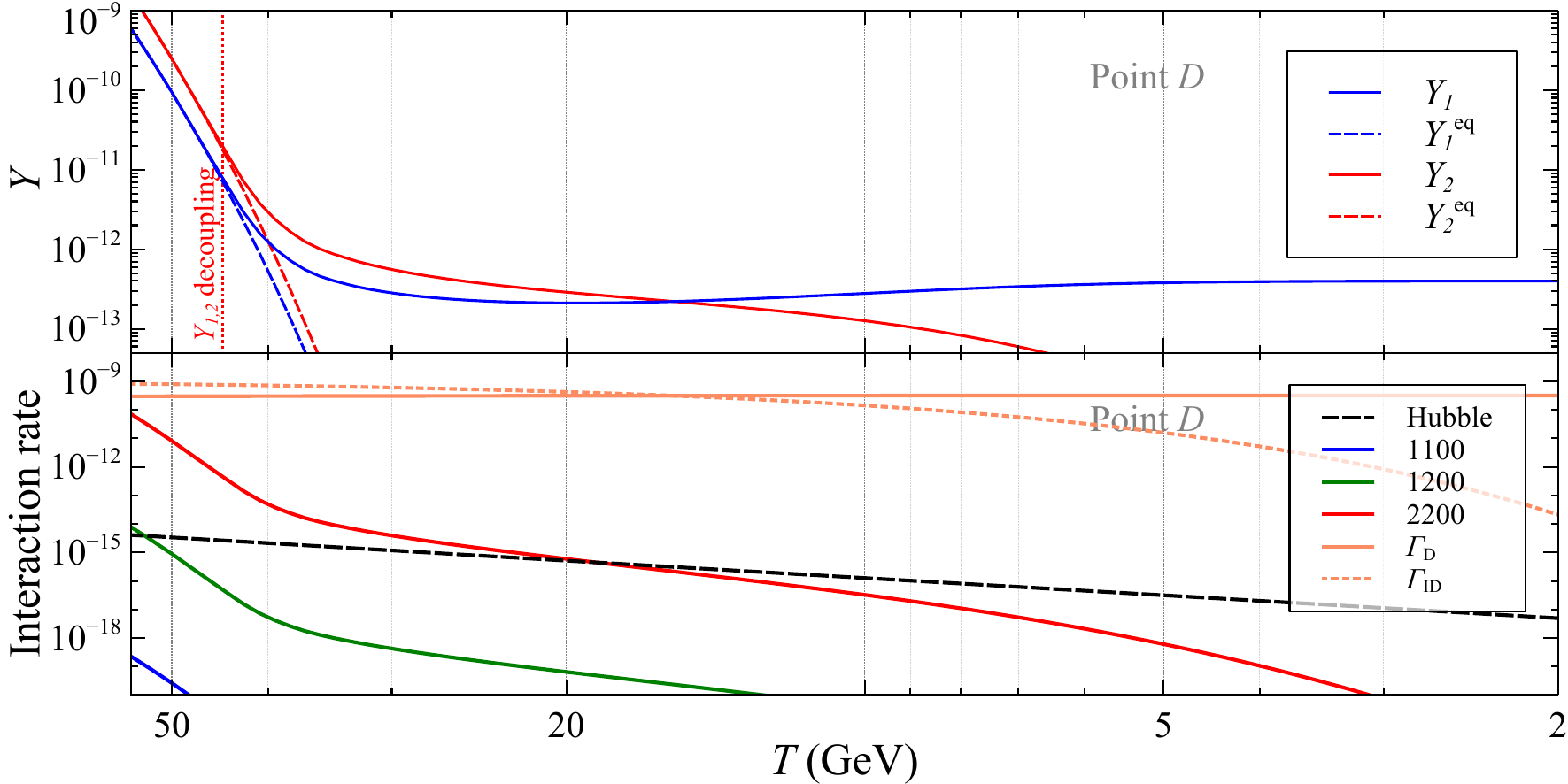}
    \caption{[\textit{Top}:] cosmological evolution of abundances of sector 1 and sector 2 particles and comparison of interaction rates with Hubble as a function of temperature for the point \textit{C} as mentioned in Table \ref{tab:tab1}. [\textit{Bottom}:] cosmological evolution of abundances of sector 1 and sector 2 particles and comparison of interaction rates with Hubble as a function of temperature for the point \textit{D} as mentioned in Table \ref{tab:tab1}.}
    \label{fig:evoplt_CD}
\end{figure*}
\begin{figure*}[tbh]
    \centering
\includegraphics[scale=0.4]{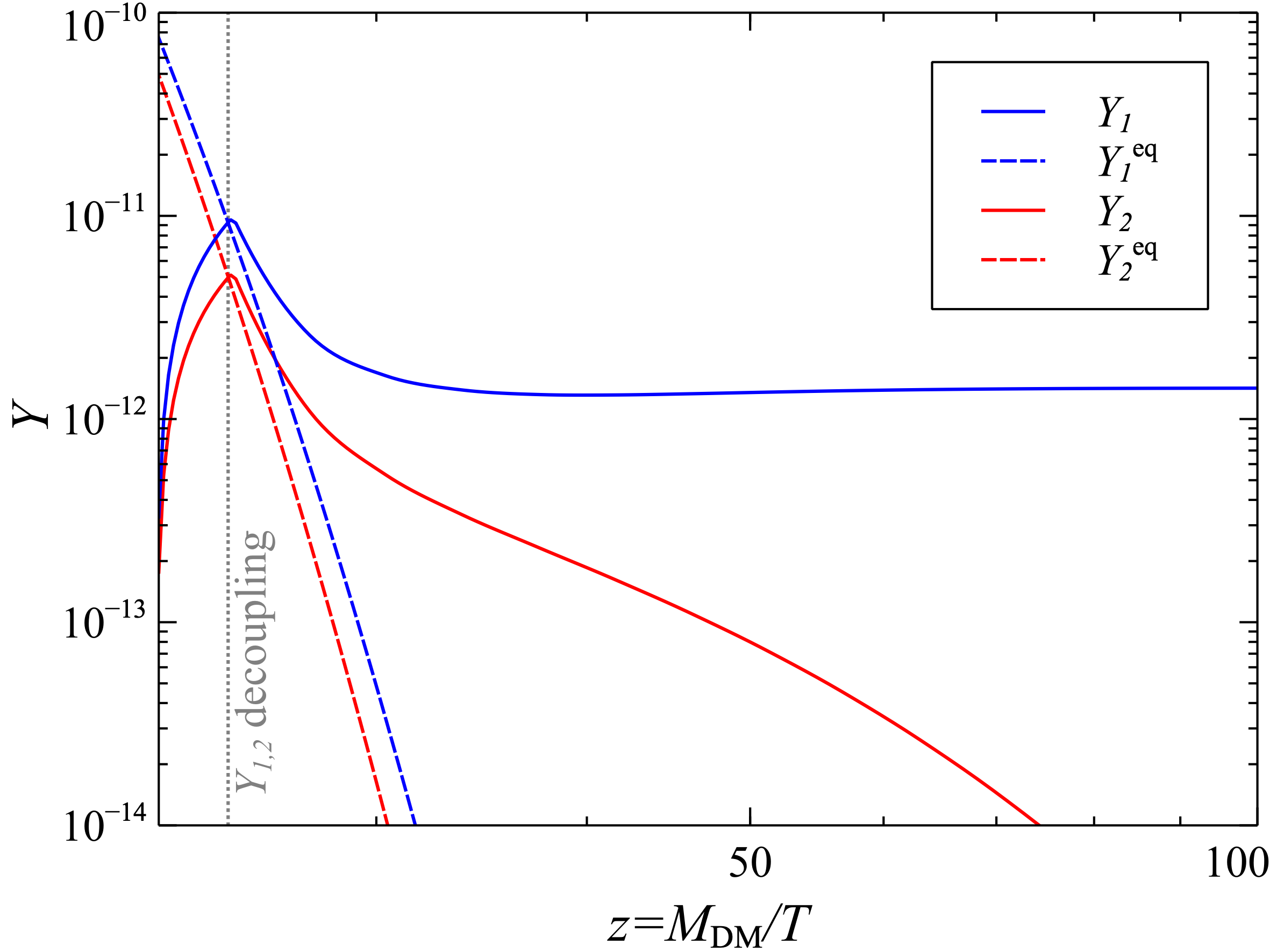}
\includegraphics[scale=0.4]{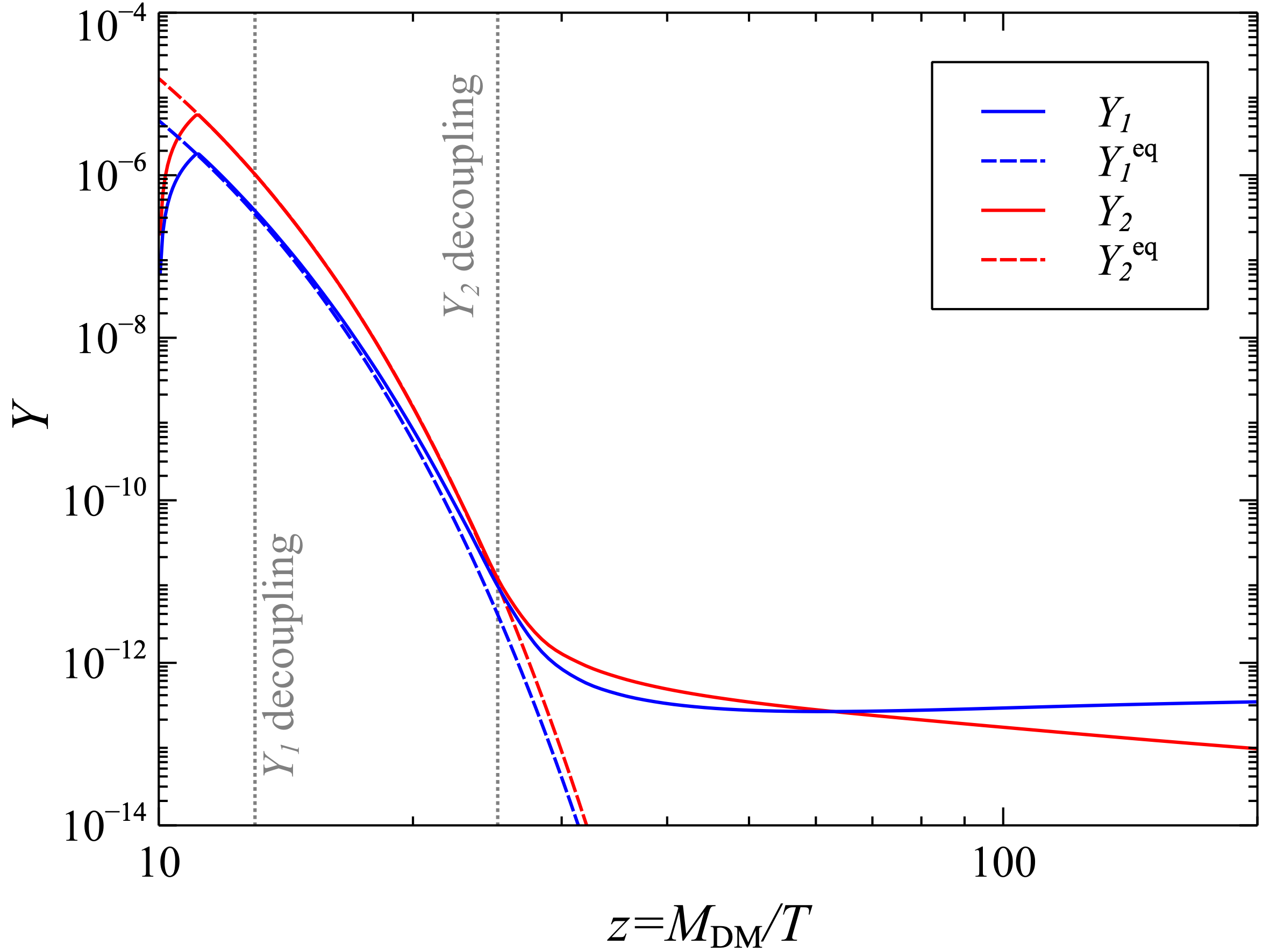}
    \caption{Cosmological evolution of abundances of sector 1 and sector 2 particles for two sets of parameters illustrating co-annihilation (\textit{left}) and co-scattering and decay effects (\textit{right}). The parameters are mentioned in Table \ref{tab:tab2} .}
    \label{fig:evoplt}
\end{figure*}
\begin{table}[tbh]
		\centering		\caption{Benchmark points for the evolution of dark sector abundances. These two points are also shown with a yellow star and a yellow diamond in Fig. \ref{fig:sinvsmdm}.}
			\begin{tabular}{|c|c|c|c|c|c|}
				\hline\hline
				Benchmark Points & $M_{\rm DM}$ (GeV) &$\Delta{M}$ (GeV) & $\sin\theta$\\
				\hline
				$BP1$&285.721&24.5826&$0.039039$\\
                \hline
				$BP2$&972.047&21.6068 &$6.97978\times10^{-7}$\\
				\hline\hline
			\end{tabular}
		\label{tab:tab2}
\end{table}
For example, for BP-$A$ ($\sin\theta=9.5\times10^{-3}$), $\Delta M=3.3$ GeV gives the correct relic abundance, as shown in the left panel of Fig. \ref{fig:delmvsmdm}. If we switch off the term $\Gamma_{2\rightarrow1}$ while solving Eqs.~(\ref{eq:Y1}) and (\ref{eq:Y2}) (i.e. without including the effect of co-scattering, decay and inverse-decay processes), the same set of parameters result in a slightly increased relic density as shown in the Table \ref{tab:tab1}. This is because the rate of co-annihilation processes decreases for small $\sin\theta$, leading to an overabundant relic that can be brought into the correct ballpark in the presence of decay and inverse-decay processes. Now in BP-$B$, we keep the $M_{\rm DM}$ and $\sin\theta$ same as BP-$A$, but with a smaller $\Delta M=1.5$ GeV. This small $\Delta M$ reduces the effect of decay and inverse-decay, and with this small $\sin\theta$, the co-annihilation effects are not sufficient, leading to an overabundant BP-$B$. However, we note that by increasing the $\sin\theta$, the relic density of BP-$B$ can be brought to the correct ballpark. In BP-$C$ ($\Delta{M}=1.3$ GeV and $\sin\theta$ same as point $A$), we consider a smaller $M_{\rm DM}=1100$ GeV, which yields a correct relic density. Similarly, we have considered another BP-$D$, with same $M_{\rm DM}$ and $\sin\theta$ as that of $C$, but with a larger $\Delta M(=22{~\rm GeV})$. We note that $D$ also gives the correct relic density. To understand why the BP-$C$ and $D$ give correct relic density, we provide two evolution plots along with the interaction rates, which are compared with the Hubble expansion rate in Fig. \ref{fig:evoplt_CD}. In the \textit{bottom} panel, we show the evolution of the abundances of dark sector particles along with the interaction rates for BP-$D$. In this case, the rate of decay/inverse decay width is larger than the Hubble expansion rate because of the large $\Delta M$. Although the co-annihilation rates (\texttt{"1200"} process) decouple early, the efficient decay/inverse decay rates keep changing the DM relic density until the \texttt{"2200"} processes go out of thermal equilibrium ($T\simeq 20$ GeV), resulting in the correct relic density. In the \textit{top} plot, we show the evolution of the abundances of dark sector particles along with the interaction rates for BP-$C$. At high temperature, the DM was in thermal equilibrium due to the efficient co-annihilation processes. In this case, the rates of decay and inverse-decay processes are weaker than those of co-annihilation processes. As a result, the DM decouples chemically from the thermal bath when the co-annihilation rates drop below the Hubble expansion rate.  However, at smaller temperatures, the inverse decay processes, which are not yet negligible, populate the abundance of sector 2 particles, leading to an enhanced rate of \texttt{"2200"} processes. Until the \texttt{"2200"} processes maintain thermal equilibrium ($T\simeq 10$ GeV) with the SM thermal bath, the decay and inverse-decay processes help to bring the DM relic density into the correct ballpark.

From the above discussion, we conclude that for $\sin\theta\gtrsim0.05$, where annihilation and co-annihilation processes are significant, the singlet and doublet decouple at the same epoch and give the correct relic density. On the other hand, for $\sin\theta\lesssim0.05$, the singlet decouples early with a larger abundance. However, it can be brought to the correct ballpark with the help of decay/inverse decay, followed by the annihilation of the doublet to SM particles. As a result, in this case, the gross features of the relic density parameter space differ significantly from the larger $\sin\theta$ ($\gtrsim0.05$) case.

So far, we have switched off the co-scattering (\texttt{"2010"} processes) effects while solving Eqs. (\ref{eq:Y1}) and (\ref{eq:Y2}). Now we solve the Eqs.~(\ref{eq:Y1}) and (\ref{eq:Y2}) without switching off the co-scattering processes and showcase the correct relic points in the plane of $M_{\rm DM}$ and $\Delta M$ in the \textit{right} panel of Fig. \ref{fig:delmvsmdm}. The various colored points represent the range of $\sin\theta$ as given in the inset of the figure. For $\sin\theta\gtrsim0.05$, the annihilation, co-annihilation, decay, and inverse-decay processes are sufficient enough to keep the DM in equilibrium with the SM bath until the sector 2 particles decouple from the SM through \texttt{"2200"} processes. Once the sector 2 particles decouple, the co-scattering processes can not alter the relic density. At most, it can interchange $Y_1$ and $Y_2$, but $Y_1+Y_2$ remains constant. This is the reason why co-scattering processes do not play any role in the large $\sin\theta$ region. For $\sin\theta\lesssim0.05$, the co-scattering processes play a significant role as the chemical decoupling of the singlet and doublet happens at different epochs.

To illustrate the effects of co-scattering and decay, we choose two sets of benchmark points as mentioned in the Table \ref{tab:tab2}. We show the cosmological evolution of the abundances of \textit{BP1} in the \textit{left} panel of Fig. \ref{fig:evoplt}. The blue (red) solid and dashed lines represent the abundance of the singlet (doublet) and its equilibrium values, respectively.  We see that the large mixing angle leads to a simultaneous decoupling of the singlet and doublet components around $z\sim24.5$. In this case, the relic is determined by the co-annihilation processes. The annihilation processes are subdominant because of the small mass splitting. We now move to \textit{BP2}. Here, we see that the singlet decouples very early, around $z\sim19$, whereas the doublet stays in equilibrium due to its gauge interactions up to $z\sim25.2$. Although the singlet decouples with a larger abundance due to co-scattering and decay processes, its relic abundance is reduced to the observed value with the help of \texttt{"2200"} processes. 
\begin{figure}[h]
    \centering   \includegraphics[scale=0.4]{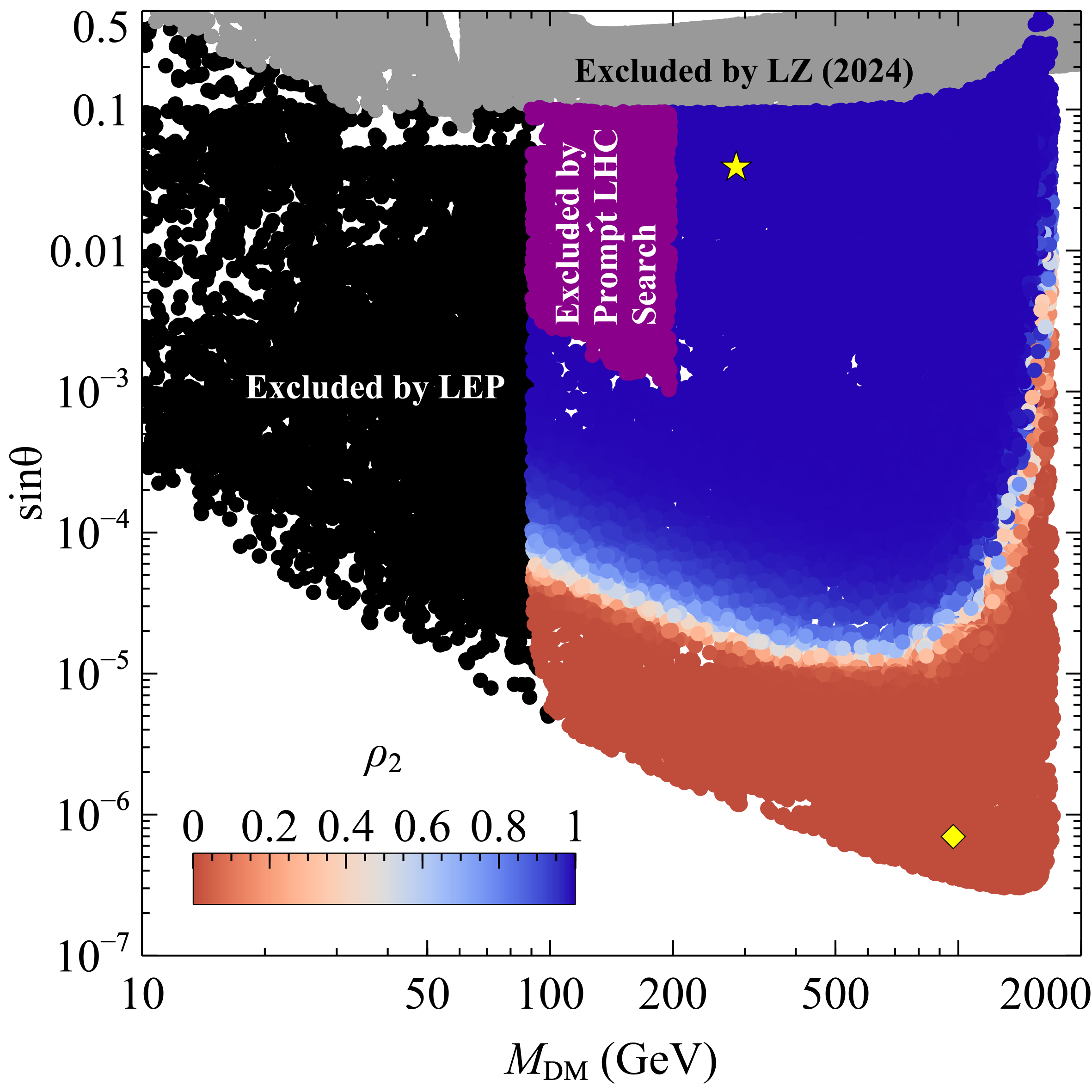}
    \caption{Correct dark matter relic points are shown in the plane of $\sin\theta$ vs $M_{\rm DM}$. The color code represents the parameter $\rho_2$. The gray shaded region is excluded by the direct detection experiment LZ \cite{LZ:2024zvo}. LEP \cite{DELPHI:2003uqw} constraint on the doublet fermion excludes the black shaded region. The magenta shaded region is excluded by the LHC prompt search \cite{ATLAS:2021moa,CMS:2021cox}.}.
    \label{fig:sinvsmdm}
\end{figure}
From the right panel of Fig. \ref{fig:delmvsmdm}, we see that co-scattering plays a significant role for $\sin\theta\lesssim0.05$. It should be mentioned that the rate of the co-scattering processes increases with a decrease in $\Delta M$. Upon adding the co-scattering processes, the co-scattering rate dominates over the decay and inverse-decay rates. As a result, the singlet will be efficiently converted to the doublet, followed by annihilation of the doublet to the SM particles (\texttt{"2200"} processes). As a result, the parameter space that previously yielded the correct relic density when co-scattering was excluded becomes under-abundant in the presence of the co-scattering processes. On the other hand, the over-abundant
points with $\sin\theta\lesssim10^{-5}$, which did not provide the correct relic density in the absence of
co-scattering processes, now yield the correct relic density once these processes are included.

To demonstrate the effect of co-scattering quantitatively, we define a new parameter,
\begin{equation}
	\rho_{2}=\frac{\Omega_{2s}h^2}{\Omega_{2s}h^2({\rm no~coscattering})},\label{eq:rho2}
\end{equation}
where $\Omega_{2s}h^2$ is obtained by solving Eqs. (\ref{eq:Y1}) and (\ref{eq:Y2}) which takes into account the effects of co-scattering in addition to all other number-changing processes, such as decay, inverse decay, annihilation, and co-annihilation. In contrast, $\Omega_{2s}h^2$ (no~co-scattering) refers to the relic density computed by solving Eqs. (\ref{eq:Y1}) and (\ref{eq:Y2}) while switching off the co-scattering processes in $\Gamma_{2\rightarrow1}$. According to the definition of $\rho_2$ provided in Eq. (\ref{eq:rho2}), a value of $\rho_2 = 1$ signifies that the relic density is independent of co-scattering effects. Consequently, any deviation from unity indicates the influence of co-scattering on the final relic density. The magnitude of this deviation serves as a measure of the impact of co-scattering; specifically, a larger deviation indicates a more significant effect. Moreover, as mentioned, $\Omega_{2s}h^2$ encompasses all relevant processes, including annihilation, co-annihilation, co-scattering, and decay, whereas $\Omega_{2s}h^2$ (no co-scattering) is limited to the processes of annihilation, co-annihilation, and decay. This distinction implies that the effective cross-section associated with $\Omega_{2s}h^2$ is always greater than or equal to that of $\Omega_{2s}h^2$ (no co-scattering). Therefore, we can conclude that: $\Omega_{2s}h^2 \le\Omega_{2s}h^2$ (no co-scattering), which leads to $\rho_2\le1$.

In Fig. \ref{fig:sinvsmdm}, we present the correct relic satisfying points in the plane of $\sin\theta-M_{\rm DM}$. The color map depicts the $\rho_2$ values. The blue colored points correspond to $\rho_2\simeq 1$. This represents the parameter space where co-scattering effects are negligible, and the relic is decided solely by annihilation and/or co-annihilation processes. As $\sin\theta$ decreases, $\rho_2$ deviates from unity, indicating the increasing significance of co-scattering processes across a wider parameter space, as shown by the light bluish and brown-colored regions. For the purpose of illustration, we show \textit{BP1} and \textit{BP2} in Fig. \ref{fig:sinvsmdm}. It is straightforward to note that even $\sin\theta$ is quite different for \textit{BP1} and \textit{BP2}, yet they can yield the correct relic density due to the significance of the different processes involved in relic computation. Below the colored region, even though DM can be produced thermally, conversion-driven processes, including decay, inverse decay, and co-scattering, are not efficient at producing the correct relic abundance, leading to an overabundant relic. If $\sin\theta$ is reduced further, the DM can never be thermalized. The relic, in this case, is determined by non-thermal processes such as freeze-in and SuperWIMP mechanisms \cite{Paul:2024prs}. The black shaded region is excluded from the large electron positron (LEP) \cite{DELPHI:2003uqw} constraint. The current direct detection limit from  LUX-ZEPLIN (LZ) \cite{LZ:2024zvo} excludes the gray region. Unlike the Dirac case, in the Majorana singlet–doublet dark matter scenario, the $Z$-mediated spin-independent direct detection channel is absent. As a result, the dominant contribution to spin-independent direct detection arises only from Higgs-mediated interactions. Since $Z$-exchange typically yields much stronger constraints, its absence significantly relaxes the bounds. Consequently, larger values of $\sin\theta$ are allowed by direct detection experiments in the Majorana case ($\sin\theta\lesssim0.45$) compared to the Dirac case ($\sin\theta\lesssim0.04$).

It is interesting to note that, in the Majorana SDDM scenario, significantly heavier dark matter masses are allowed. In the Majorana case, the effective cross section is enhanced due to additional contributing channels. This allows the observed relic density to be satisfied at higher dark matter masses than in the Dirac scenario. In particular, masses up to about 1750 GeV remain viable, whereas in the Dirac case the allowed mass range is restricted to approximately 750 GeV.
\begin{figure}[h]
    \centering   \includegraphics[scale=0.4]{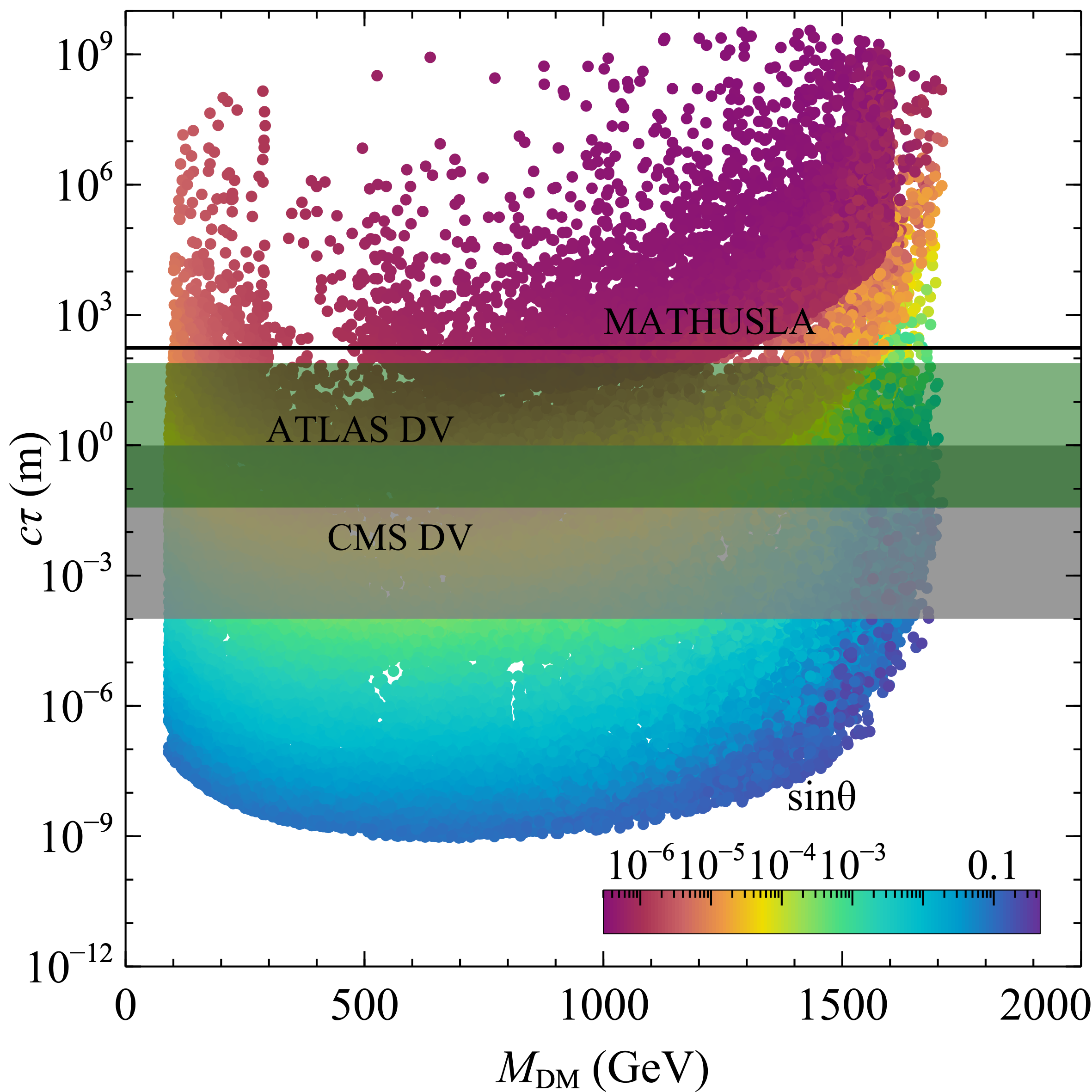}
    \caption{Decay length of the doublet fermion as a function of DM mass with $\sin\theta$ in the color code. The gray and green shaded regions represent the sensitivities of CMS \cite{CMS:2024trg} and ATLAS \cite{ATLAS:2022gbw}, respectively, and the MATHUSLA \cite{MATHUSLA:2019qpy} sensitivity is shown with a black solid line.}
    \label{fig:DV}
\end{figure}

\subsection{Collider searches}

\subsubsection{Displaced vertex signatures}
We note that the doublet fermion ($\psi^\pm$) can be produced because of
its gauge interaction, which further can decay to DM and charged leptons via the singlet-doublet mixing through off-shell gauge boson exchange. This can give rise to displaced vertex signatures in the LHC \cite{ATLAS:2022gbw,CMS:2024trg} and, in the future,
MATHUSLA \cite{MATHUSLA:2019qpy}. As already discussed, conversion-driven processes allow smaller $\sin\theta(\lesssim\mathcal{O}(10^{-3}))$ which results in smaller decay widths of  $\psi^\pm$, thereby increasing the decay length, which can fall in the reach of LHC and MATHUSLA \cite{Paul:2024prs}. 
\begin{figure}[h]
    \centering   \includegraphics[scale=0.4]{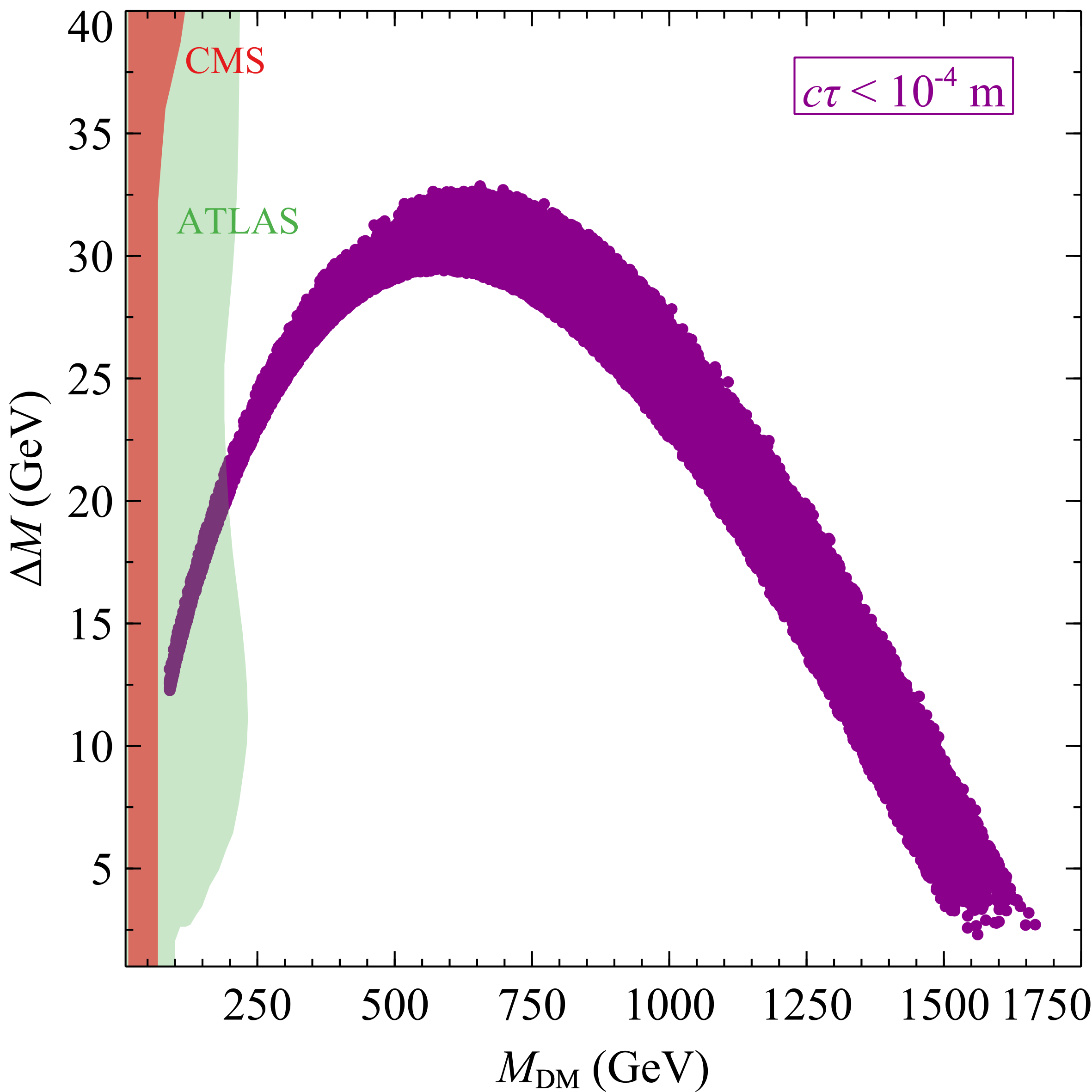}
    \caption{Prompt decay parameter space ($c\tau<10^{-4}$ m) is shown  in the plane of $\Delta{M}$ vs $M_{\rm DM}$. The 95\% exclusion limits from CMS \cite{CMS:2021cox} and ATLAS \cite{ATLAS:2021moa} using $pp\rightarrow 3l+{E^{\rm miss}_T}$ are shown with red and green shaded regions, respectively.}
    \label{fig:delMvsmdm_prompt}
\end{figure}
We calculate the decay length ($c\tau$) for the points satisfying DM relic density, direct detection, and LEP constraints and show them in the plane of $c\tau$ vs $M_{\rm DM}$ in Fig. \ref{fig:DV} considering the decay mode $\psi^\pm\rightarrow\chi_3 e^\pm\nu_e$. The color code depicts the values of $\sin\theta$. The region of sensitivity of the CMS ($10^{-4}{~\rm m}-1{~\rm m}$) and ATLAS ($0.04{~\rm m}-72.4{~\rm m}$) displaced vertex searches is shown with gray and green shaded regions, respectively. The figure shows that the inclusion of co-scattering processes increases the lifetime of the dark partners, bringing them within the sensitivity reach of the LHC and MATHUSLA.

\subsubsection{Prompt LHC search} If the decay length of the charged doublet is smaller than $10^{-4}$ m, it decays before reaching the detector volume, leading to prompt decay signatures at the LHC. From Fig.~\ref{fig:DV}, we select all points with $c\tau < 10^{-4}$ m, which correspond to $3.4\times 10^{-4} < \sin\theta<0.4$, that are displayed in the $\Delta M$–$M_{\rm DM}$ plane in Fig.~\ref{fig:delMvsmdm_prompt}. The exclusion limits from CMS \cite{CMS:2021cox} and ATLAS \cite{ATLAS:2021moa} using $pp\rightarrow 3l+{E^{\rm miss}_T}$ (${E^{\rm miss}_T}$ denotes the missing transverse energy) are overlaid as red and green shaded regions, respectively. We find that dark matter masses $M_{\rm DM} \lesssim 197$ GeV with $c\tau < 10^{-4}$ m and $\sin\theta\gtrsim 10^{-3}$ are excluded by prompt searches at ATLAS. In Fig.~\ref{fig:sinvsmdm}, the excluded regions from prompt LHC searches are shown as a magenta shaded region in the $\sin\theta$–$M_{\rm DM}$ plane.

\begin{figure}[h]
    \centering   \includegraphics[scale=0.4]{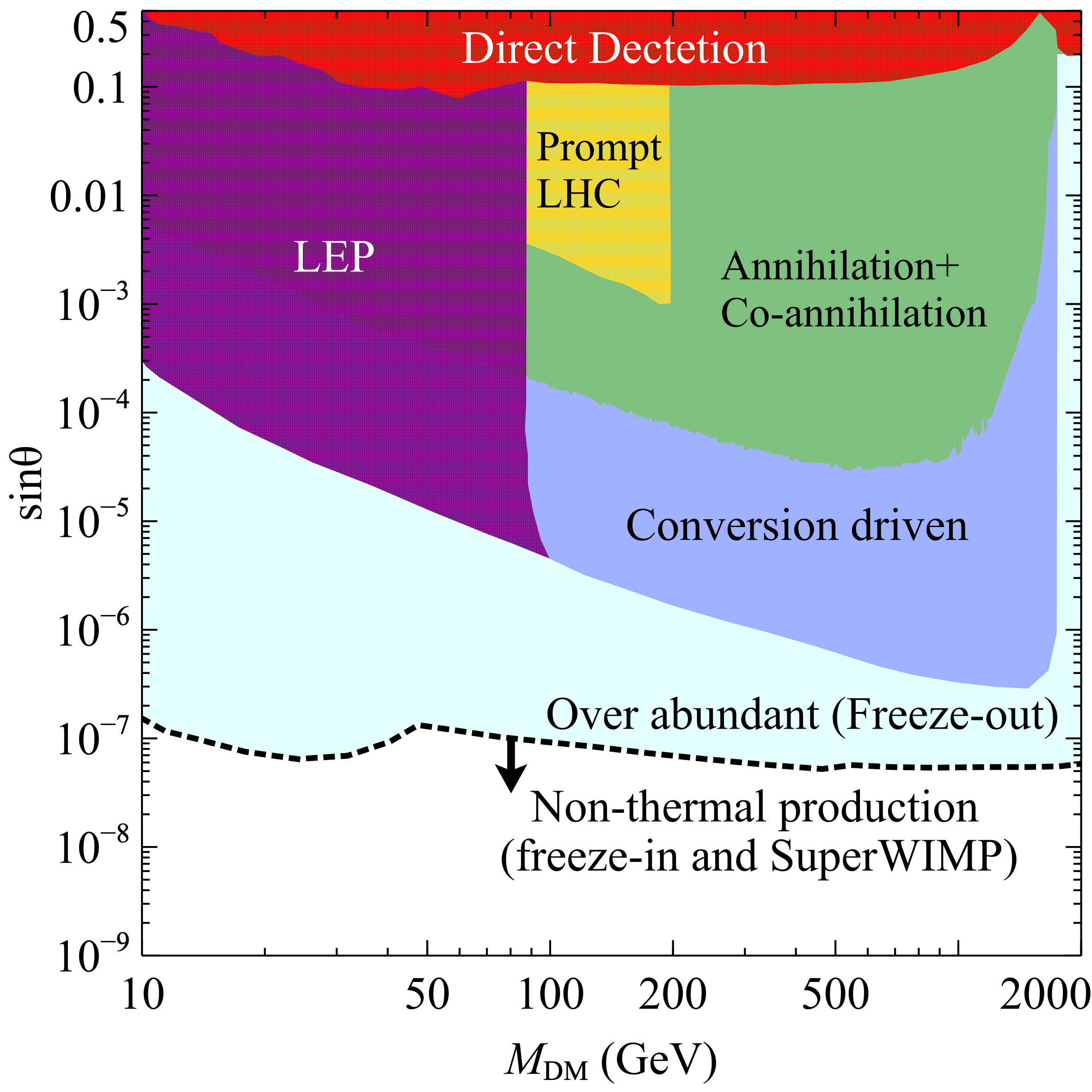}
    \caption{Phase diagram illustrating the SDDM parameter space. The black dashed line separates the thermal regime (above) from the non-thermal regime (below).}
    \label{fig:summ}
\end{figure}
\section{Conclusion and future outlook}\label{sec:conc}

In this work, we analyzed the parameter space of the singlet-doublet Majorana DM considering the thermal freeze-out scenario. We found that the combined upper limit on the mixing angle from  LZ and LEP leads to  $\sin\theta\lesssim0.45$. We summarize our analysis in Fig. \ref{fig:summ}, in the plane of $M_{\rm DM}$ and $\sin\theta$. In the green-shaded region, annihilation and co-annihilation processes determine the relic abundance of DM depending on the value of $\Delta{M}$. In the blue-shaded region, conversion-driven processes determine the relic. The LEP exclusion is shown with the magenta shaded region. The exclusion from the LHC prompt search is shown in the yellow-shaded region. In the cyan shaded region, even though the DM can be in thermal equilibrium, due to such a small $\sin\theta$, it decouples early with a larger abundance. Conversion-driven processes are inefficient in this region. The black dashed line serves as the boundary, above which DM can be brought to thermal equilibrium with the SM thermal bath. In the region below the black dashed line, the DM never reaches thermal equilibrium because of the small SD mixing angle. We find the allowed ranges of DM mass and $\sin\theta$ for the thermal relic of DM to be $100~{\rm GeV}\lesssim M_{\rm DM}\lesssim1750$ GeV and $2\times10^{-7}\lesssim\sin\theta\lesssim0.45$ for $\Delta{M}>1$ GeV. We also note that incorporating conversion-driven processes substantially enlarges the region of parameter space accessible to both the LHC and MATHUSLA experiments.

\section*{Acknowledgements}
P. K. P. would like to acknowledge the Ministry of Education, Government of India, for providing financial support for his research via the Prime Minister’s Research Fellowship (PMRF) scheme.

\appendix
\section{Fermion mass diagonalizaton}\label{app:massdiag}

The neutral fermion mass matrix can be written in the basis: $((\psi^0_R)^c,\psi^0_L,(\chi)^c)^T$ as
\begin{equation}
\mathcal{M}=\begin{pmatrix}
	0 & M_\Psi & \frac{m_D}{\sqrt{2}}\\
	M_\Psi&0&  \frac{m_D}{\sqrt{2}}\\
	 \frac{m_D}{\sqrt{2}} &  \frac{m_D}{\sqrt{2}}& M_\chi
\end{pmatrix},
\end{equation}
where $m_D=y_\chi v_h/\sqrt{2}$, and $v_h$ is the vev of the SM Higgs boson, $H$.
The mass matrix can be diagonalized with a unitary matrix of the form
\begin{eqnarray}
U(\theta)&=&P.U_{13}(\theta_{13}=\theta).U_{23}(\theta_{23}=0).U_{12}(\theta_{12}=\frac{\pi}{4})\nonumber\\&=&\begin{pmatrix}
     1&0&0\\
     0&e^{i\pi/2}&0\\
     0&0&1
 \end{pmatrix}.\begin{pmatrix}
     \cos\theta/\sqrt{2}&\cos\theta/\sqrt{2}&\sin\theta\\
     -1/\sqrt{2}& 1/\sqrt{2}&0\\
    - \sin\theta/\sqrt{2}& - \sin\theta/\sqrt{2} &\cos\theta/\sqrt{2}
\end{pmatrix}\nonumber\\&=&\begin{pmatrix}
     \cos\theta/\sqrt{2}&\cos\theta/\sqrt{2}&\sin\theta\\
     -i/\sqrt{2}& i/\sqrt{2}&0\\
    - \sin\theta/\sqrt{2}& - \sin\theta/\sqrt{2} &\cos\theta/\sqrt{2}
 \end{pmatrix}
\end{eqnarray}
The diagonal mass matrix is given by 
\begin{eqnarray}
    \mathcal{M}_{\rm diag}=U.\mathcal{M}.U^T,
\end{eqnarray}
with eigenvalues $M_{\chi_1},M_{\chi_2}$, and $M_{\chi_3}$. Then the flavor masses can be expressed in terms of the physical masses as
\begin{align}\label{eq:mpsimchi}
    M_\Psi&=\frac{1}{2}\left( M_{\chi_1}+M_{\chi_3}+(M_{\chi_1}-M_{\chi_3})\sec2\theta-2m_D\tan2\theta \right)\nonumber\\ M_\chi&=\frac{1}{2}\left( M_{\chi_1}+M_{\chi_3}-(M_{\chi_1}-M_{\chi_3})\sec2\theta+2m_D\tan2\theta \right),
\end{align}
where the mixing angle is given as
\begin{eqnarray}  \tan2\theta=\frac{2m_D}{M_\Psi-M_\chi} \label{eq:tan2th}
\end{eqnarray}
Now using Eq \ref{eq:mpsimchi}, Eq.~\ref{eq:tan2th} simplifies to
\begin{eqnarray}\label{eq:sin2th}  \sin2\theta=\frac{2m_D}{M_{\chi_1}-M_{\chi_3}}.
\end{eqnarray}

\section{Lagrangian in mass basis}\label{app:lagMass}
The interaction Lagrangian between $\chi_1,\chi_2,\chi_3$ and $\psi^-$ with the SM gauge boson is given by
\begin{eqnarray}
    \mathcal{L}_{\rm Gauge}&=&\left(\frac{e}{2s_wc_w}\right)(-c_{\theta}\overline{\chi_1}i\gamma^\mu Z_\mu P_L\chi_2-s_{\theta}\overline{\chi_2}i\gamma^\mu Z_\mu P_L\chi_3\nonumber\\&+&{\rm H.c.})+\frac{e}{\sqrt{2}s_w}( c_{\theta}\overline{\chi_1}\gamma^\mu W^+_\mu\psi^-+c_{\theta}\overline{\chi_2}\gamma^\mu W^+_\mu\psi^-\nonumber\\&-&s_{\theta}\overline{\chi_3}\gamma^\mu W^+_\mu\psi^- )+\frac{e}{\sqrt{2}s_w}( c_{\theta}\psi^+\gamma^\mu W^-_\mu\chi_1\nonumber\\&+&c_{\theta}\psi^+\gamma^\mu W^-_\mu\chi_2-s_{\theta}\psi^+\gamma^\mu W^-_\mu\chi_3 )-
    e\psi^+\gamma^\mu A_\mu\psi^-\nonumber\\&-&\left( \frac{ec_{2w}}{2s_wc_w} \right) \psi^+\gamma^\mu Z_\mu  \psi^-, 
\end{eqnarray}
where $s_w=\sin\theta_w,c_w=\cos\theta_w,c_{2w}=\cos2\theta_w,s_\theta=\sin\theta,c_\theta=\cos\theta$. The Yukawa Lagrangian can be written as
\begin{eqnarray}
    \mathcal{L}_{\rm Yukawa}&=&\frac{y_\chi}{\sqrt{2}}\left( \sin2\theta\left(\overline{\chi_1}h\chi_1-\overline{\chi_3}h\chi_3\right)\right.\nonumber\\&&\left.+\cos2\theta \left(\overline{\chi_1}h\chi_3-\overline{\chi_3}h\chi_1\right)  \right)
\end{eqnarray}

\section{Feynman diagrams}\label{app:FD}
\subsection*{Annihilation processes of DM}
\begin{figure}[H]
    \centering
    \includegraphics[scale=0.5]{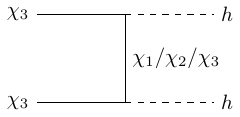}
    \includegraphics[scale=0.5]{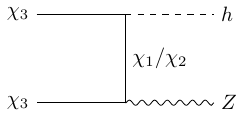}
    \includegraphics[scale=0.5]{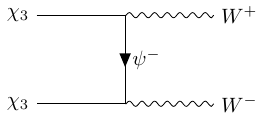}
    \includegraphics[scale=0.5]{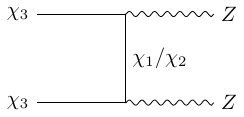}
    \includegraphics[scale=0.4]{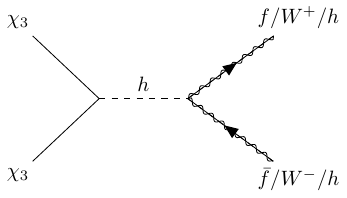}
    \caption{DM annihilating to the SM particles through 1100 processes.}
    \label{fig:annDM}
\end{figure}

Here, we have provided the Feynman diagram of annihilation (corresponding to 1100 processes) of sector 1 particle (i.e., DM or $\chi_3$) to sector 0 particles (i.e., SM particles) in Fig. \ref{fig:annDM}.

\subsection*{Co-annihilation between sector-1 and sector-2 particles}
\begin{figure}[H]
    \centering
    \includegraphics[scale=0.4]{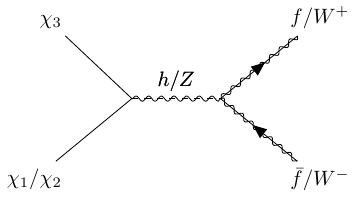}
    \includegraphics[scale=0.4]{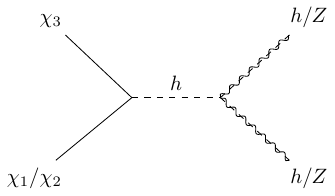}
    \includegraphics[scale=0.4]{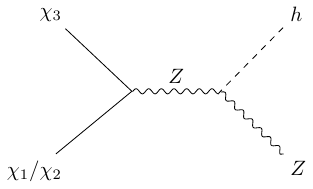}
    \includegraphics[scale=0.5]{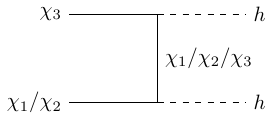}
    \includegraphics[scale=0.5]{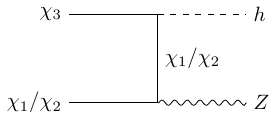}
    \includegraphics[scale=0.5]{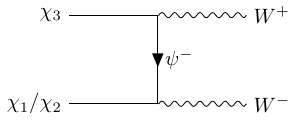}
    \includegraphics[scale=0.5]{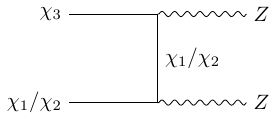}
    \includegraphics[scale=0.5]{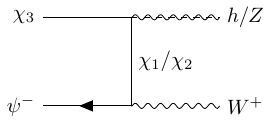}
    \includegraphics[scale=0.5]{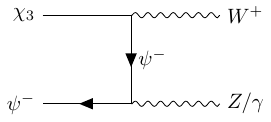}
    \includegraphics[scale=0.4]{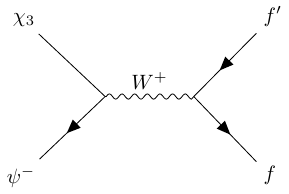}
    \includegraphics[scale=0.4]{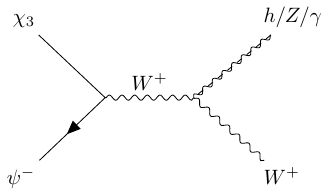}
    \caption{DM co-annihilating with the doublet components to the SM particles through 1200 processes.}
    \label{fig:coannDM1}
\end{figure}

The co-annihilation (corresponds to 1200 processes) between the sector 1 ($\chi_3$) and sector 2 particles (i.e. $\chi_1,\chi_2,\psi^{-}$ ) to SM particles are shown in Fig. \ref{fig:coannDM1}.
\subsection*{Annihilation and co-annihilation among the sector 2 particles}
\begin{figure}[H]
    \centering
    \includegraphics[scale=0.5]{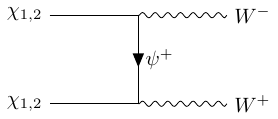}
    \includegraphics[scale=0.5]{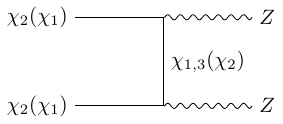}
    \includegraphics[scale=0.4]{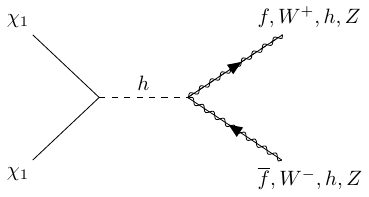}
    \includegraphics[scale=0.5]{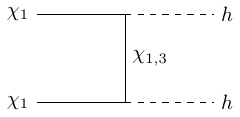}
    \includegraphics[scale=0.5]{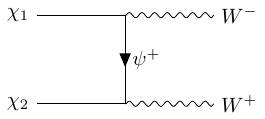}
    \includegraphics[scale=0.4]{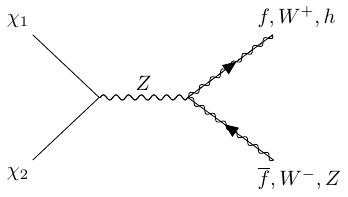}
    \includegraphics[scale=0.5]{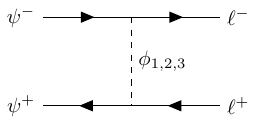}
    \includegraphics[scale=0.5]{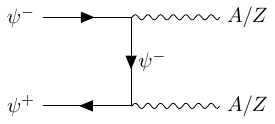}
    \includegraphics[scale=0.4]{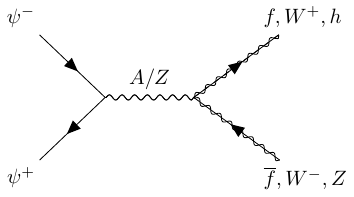}
    \includegraphics[scale=0.5]{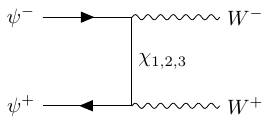}
    \caption{Annihilation and co-annihilation among the sector 2 particles to SM particles corresponding to the 2200 processes.}
    \label{fig:annDM2}
\end{figure}

Annihilation and co-annihilation (corresponding to 2200 processes) among the sector 2 particles to SM particles are given in Fig. \ref{fig:annDM2}.
\subsection*{Co-scattering between DM and doublet fermion}
\begin{figure}[h]
    \centering
    \includegraphics[scale=0.4]{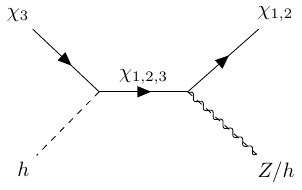}
    \includegraphics[scale=0.4]{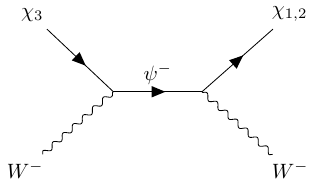}
    \includegraphics[scale=0.4]{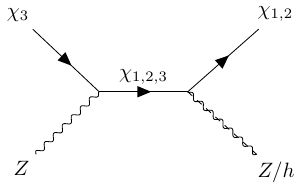}
    \includegraphics[scale=0.5]{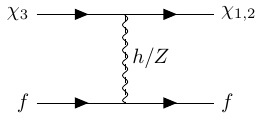}
    \includegraphics[scale=0.5]{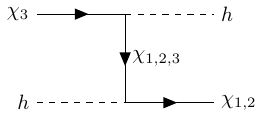}
    \includegraphics[scale=0.5]{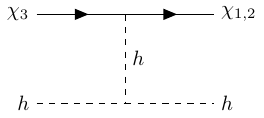}
    \includegraphics[scale=0.5]{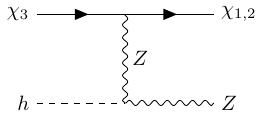}
    \includegraphics[scale=0.5]{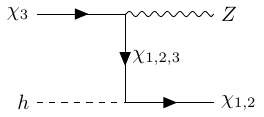}
    \includegraphics[scale=0.5]{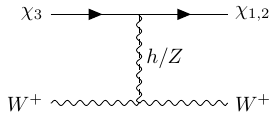}
    \includegraphics[scale=0.5]{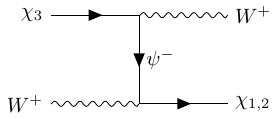}
    \includegraphics[scale=0.5]{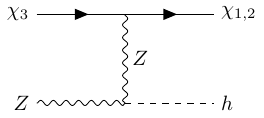}
    \includegraphics[scale=0.5]{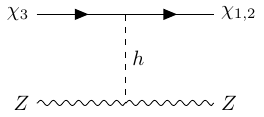}
    \includegraphics[scale=0.5]{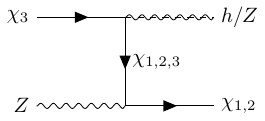}
    \includegraphics[scale=0.5]{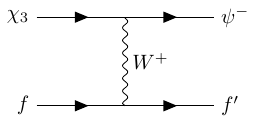}
    \includegraphics[scale=0.5]{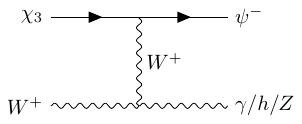}
    \includegraphics[scale=0.5]{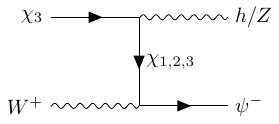}
    \includegraphics[scale=0.5]{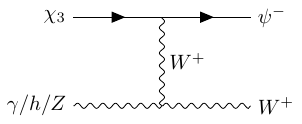}
    \includegraphics[scale=0.5]{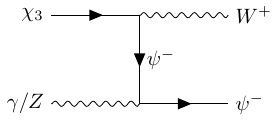}
    \includegraphics[scale=0.4]{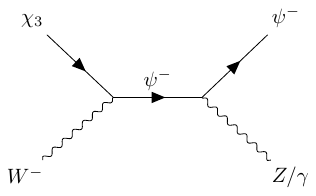}
    \includegraphics[scale=0.4]{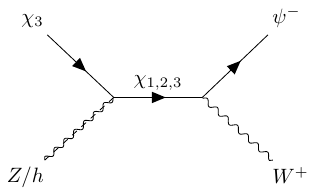}
    \caption{DM co-scattering with the SM bath particles to the doublet states through \texttt{2010} processes.}
    \label{fig:cos-catt1}
\end{figure}

Co-scattering (corresponding to \texttt{2010} processes) among the sector 1 particle and sector 2 particles are given in Fig. \ref{fig:cos-catt1}.

\subsection*{DM production from decay}
\begin{figure}[h]
    \centering
    \includegraphics[scale=0.3]{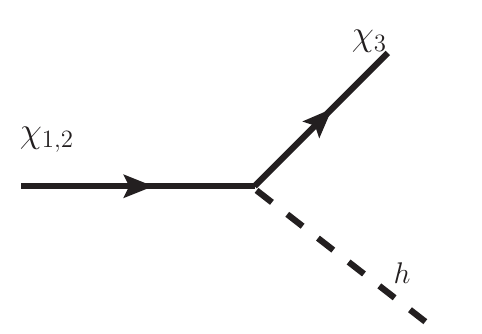}
    \includegraphics[scale=0.3]{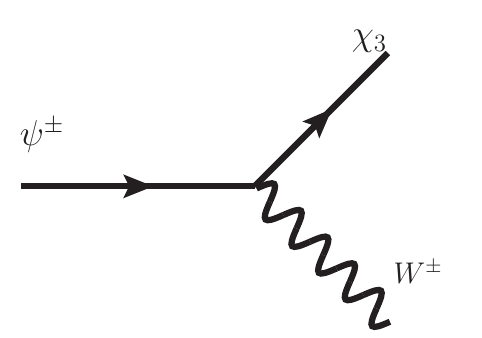}
    \includegraphics[scale=0.3]{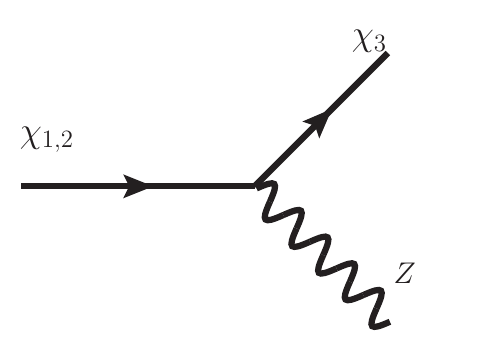}
    \includegraphics[scale=0.3]{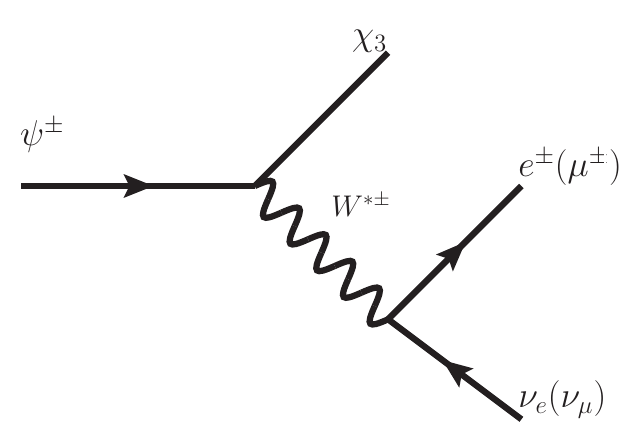}
    \caption{DM production from doublet fermion decay.}
    \label{fig:decay}
\end{figure}

We list the two and three-body decays of the doublet fermion for the production of DM in Fig. \ref{fig:decay}.

\section{Correct DM relic parameter space applying direct detection and LEP constraints}
\begin{figure}[H]
    \centering
    \includegraphics[scale=0.4]{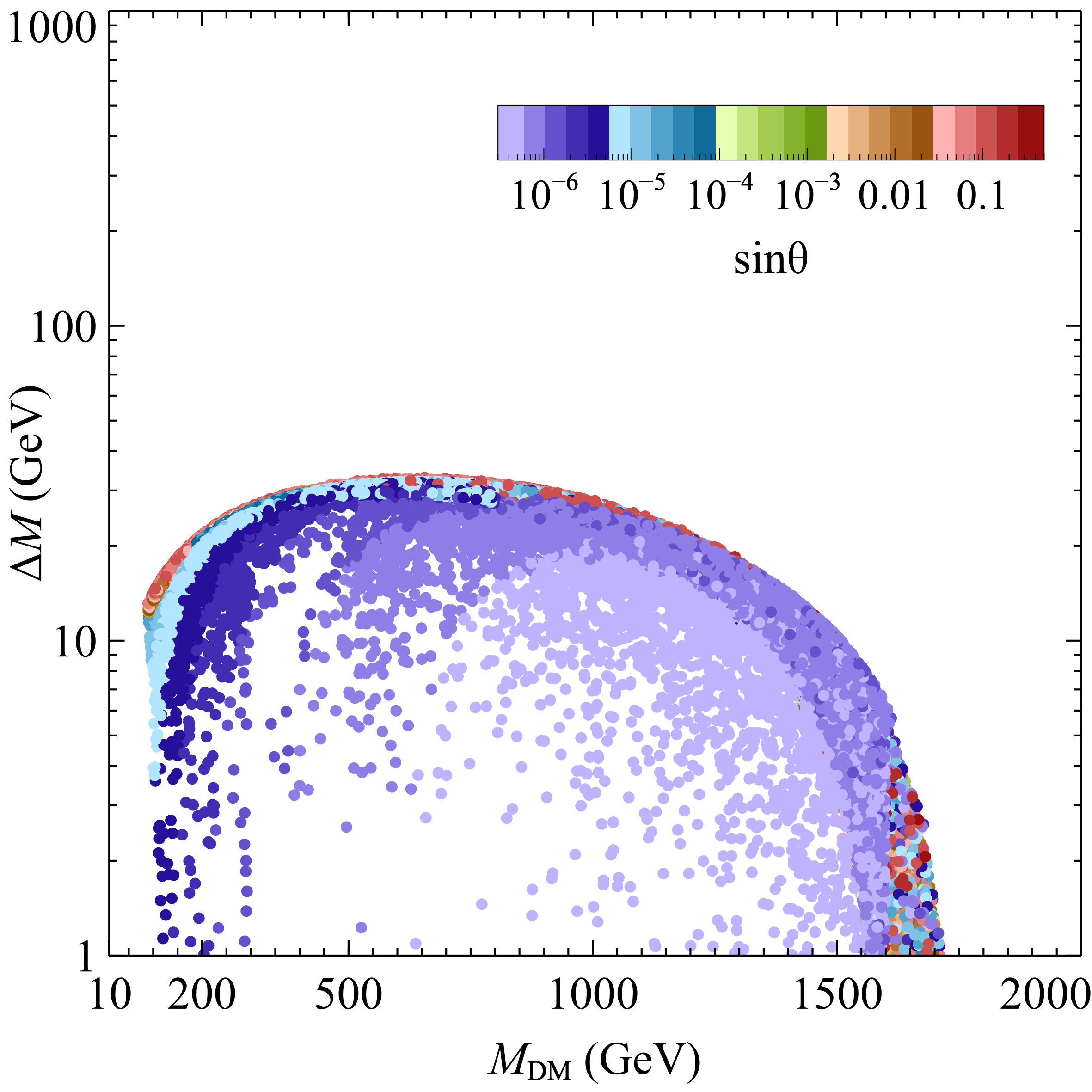}
    \caption{The correct relic points allowed by LEP, prompt LHC searches, and direct detection bound. The color band represents the $\sin\theta$ values.}
    \label{fig:delmvsmdm2}
\end{figure}
We show the correct relic points allowed from LEP and direct detection constraints in the plane of $M_{\rm DM}$ and $\Delta M$ in Fig. \ref{fig:delmvsmdm2}. The color band represents the $\sin\theta$ values.

%

\end{document}